\documentclass[prb,aps,twocolumn,floatfix]{revtex4}
\usepackage{graphicx}
\usepackage{bm}
\begin{document}
\title{Thermodynamics of the two--dimensional Falicov--Kimball model: 
a~classical Monte~Carlo study}
\author{ Maciej M. Ma\'ska}
\email{maciek@phys.us.edu.pl}
\author{Katarzyna Czajka}
\affiliation{
Department of Theoretical Physics, Institute of Physics,
University of Silesia, 40-007 Katowice,
Poland}

\date{\today}

\begin{abstract}
The two--dimensional Falicov--Kimball (FK) model is analyzed using Monte
Carlo method. In the case of concentrations of both itinerant and localized
particles equal to 0.5 we determine temperature dependence of
specific heat, charge density wave susceptibility and density--density
correlation function. In the weak interaction regime we find a first order
transition to the ordered state and anomalous temperature dependence of the
correlation function. We construct the phase diagram of half--filled FK model.
Also, the role of next--nearest--neighbor hopping on the phase diagram is analyzed.
Lastly, we discuss the density of states and the spectral functions for the
mobile particles in weak and strong interaction regime.
\end{abstract}

\pacs{71.10.-w,71.10.Fd,71.30.+h}

\maketitle

\section{Introduction}

The study of correlated electron systems has attracted great
interest over the last decades. Much of this effort was
devoted to simple Hamiltonians that may contain the basic
interactions to explain some properties of these systems.
A common point of departure for many theoretical studies is
the Hubbard model\cite{Hubbard} which had been proposed to
describe electron correlations in the narrow--band systems. This
simple model has been extensively investigated over the past
forty years, mostly in connection with the metal--insulator
transition. After discovery of high temperature
superconductors,\cite{HTSC} it was argued\cite{Anderson} that
the same simple model could possibly capture some of the
physics of these materials.

The Hubbard model is represented by the following Hamiltonian:
\begin{equation}
{\cal H}_{\rm Hubb}= \sum_{ij\sigma} t_{ij}c^\dagger_{i\sigma}c_{j\sigma}
+U\sum_i n_{i\uparrow} n_{i\downarrow},
\label{hubbard}
\end{equation}
where $c^\dagger_{i\sigma}$ ($c_{i\sigma}$) creates (annihilates) a conduction
electron with spin $\sigma$ at lattice site $i$. The hopping integral $t_{ij}$
is usually assumed to be non-zero for nearest neighboring sites $i$ and $j$
only. $U$ is the on--site Coulomb interaction.
Although at first sight such a model may seem oversimplified,
only few rigorous results are known. They include
the one--dimensional solution through the Bethe ansatz,\cite{Lieb-Wu}
the Nagaoka's theorem\cite{Nagaoka} and some statements which become
exact in infinity dimensions.\cite{G-K,Jar} All other results are approximate
(mostly of mean--field or perturbative type) or obtained for finite
lattices, mainly by Lancz\"os or quantum Monte Carlo (MC) calculations.

Therefore, our starting point is a simpler model, that can be viewed as
a limiting case of a generalized (asymmetric) Hubbard model. In the
original Hubbard Hamiltonian (\ref{hubbard}) the hopping integral $t$
is spin independent. However, one can assume that the mass of spin--up
and spin--down electrons are different, and in the limit of infinitely
massive spin--down electrons, they localize and only spin--up ones occur
in the first sum in Eq. (\ref{hubbard}). Such an approximation to the
Hamiltonian (\ref{hubbard}) was already used by Hubbard.\cite{Hubbard}

Denoting $c^\dagger_{i\uparrow}$ ($c_{i\uparrow}$)
by $c^\dagger_i$ ($c_i$), $n_{i\uparrow}$ by $n_i$ and
$n_{i\downarrow}$ by $w_i$, the resulting Hamiltonian reads
\begin{equation}
{\cal H}=-t\sum_{\langle ij\rangle}c^\dagger_i c_j + U \sum_i n_i w_i.
\label{FK}
\end{equation}
Here, $w_i$ is equal to 0 or 1, according to whether the site $i$ is
occupied or unoccupied by a massive particle. The Hamiltonian
(\ref{FK}) is known as the FK Hamiltonian.\cite{FK}
Within the framework of a common interpretation of the FK
Hamiltonian, there are two species of particles: itinerant
electrons and classical localized particles. The classical particles
have various physical interpretations: localized ($f$) electrons,
spin--down electrons, ions, impurities, nucleons. In the following we
refer to them as ``ions''. The ions interact on--site with electrons.
There are no direct interactions neither between the electrons
nor between the ions. However, the electron--ion Coulomb interaction
leads to an effective interaction between ions. As a result, for
a given number of ions, the ground state energy depends on their
distribution.

The FK model has a long and successful history in
dealing with correlated electron systems. Introduced in 1969
to describe the metal--semiconductor transition in SmB$_6$ and
related materials,\cite{FK} has been also studied as a model of
crystallization due to effective interactions mediated by band
electrons,\cite{Ken-Lieb1,Ken-Lieb2} as a binary alloy model and many others.
The FK model is also useful for describing systems that exhibit a phase
separation\cite{FL,FLU,sep1,sep2,sep3} and stripe formation.\cite{LFB,str1,str2}

The FK Hamiltonian is over thirty five years old, or even older if one takes
into account that Hubbard used it in 1963 as an approximation to his
model. However, while it is simpler than the Hubbard model, the general
solution is also not known. On the other hand, there is much more
rigorous results for the FK model, then for the Hubbard one.
One of the most important, proved by Kennedy and Lieb,\cite{Ken-Lieb1,Ken-Lieb2}
says  that at low enough temperature the half--filled Falicov--Kimball model
possesses a long range order, i.e., the ions form a checkerboard
pattern, the same as in the ground state. It is a phase, where the
lattice can be divided into two interpenetrating sublattices A and
B in such a way, that all nearest neighbors of a site from
sublattice A belongs to sublattice B and {\em vice versa} and ions
occupy only one of them. This result holds
for arbitrary bipartite lattices in dimensions $d\ge 2$ and for all
values of the interaction strength $U$. Apart from this result not very much is
known about solutions for two--dimensional systems.\cite{LFB,2D,Raedt}
Most of other results for the FK model concern one--dimension case
\cite{1Da,1Db,1Dc,1Dd} or the infinite--dimensional limit.\cite{infD1,infD2} 
The present interest in this model was renewed in connection with
developing new calculation methods that can be applied to the
FK model. In particular, the dynamical mean--field theory\cite{MetzVol}
(DMFT) provides the exact solution of the FK model in infinite
dimensions.\cite{infD1,infD2} While it is known that the DMFT captures
many key features of the FK model even in finite dimensions, this approach
also has some limitations. The local approximation used in the DMFT
does not allow to incorporate any nonlocal correlations, which are
necessary to describe many of phases that are expected to realize
in the FK model. An extension to the DMFT that include short--range
dynamical correlations has been recently proposed.\cite{DCA,DCA1}
Within this approach, called dynamical cluster approximation (DCA),
the ``impurity'' used in the DMFT is replaced by a finite--size cluster.

At non--zero temperature the partition function has to be
calculated for each ionic configuration, and then one has to average
over the configurations (``annealing''). However, since the number
of ionic configurations
increases very rapidly with the size of the system, such a procedure
is possible in some cases only. One of them is called ``restricted phase
diagram'' method,\cite{2D} where only periodic configurations on infinite
lattice are taken into account. In order to include {\em all} possible
configurations, one has to significantly restrict the size of the lattice.
In such a case an exact diagonalization technique can be applied. \cite{farkas}
This approach is particularly useful for low--dimensional systems, where
the finite size effects are of relatively small importance. Moreover, for
linear chains it is possible to perform $N\rightarrow\infty$ extrapolation
in a systematic way. On the other hand, it is difficult to perform the exact 
diagonalization studies at finite temperatures. In order to calculate the partition 
function one has to run a summation over $2^N$ ionic configurations. Therefore, 
the maximum size of the clusters suitable for exact diagonalization study
is strongly limited. In Ref. \onlinecite{farkas} the largest clusters consisted
of 24 lattice sites.

In the present contribution we propose to avoid these limitations by using the
MC method, where thermodynamic quantities for the FK model are determined
by sampling the ionic configuration space stochastically. This approach allows
us to investigate clusters of up to a few hundred sites.
Since the frozen particles are classical ones, we do not have to use
the quantum MC algorithm with the ``fermionic sign'' problem, and
thus the calculations are not restricted to the high--temperature regime.

The present calculations are restricted to the neutral half--filled case,
where the number of ions is equal to the number of electrons and their sum
is equal to the number of lattice sites. The developed formalism can be
straightforwardly used away from half--filling, however, the computational
effort in this case is much larger, mainly due to the richness of the
zero-temperature phase diagram of the FK model. Nevertheless, some results
for various concentrations have already been obtained.\cite{pss}

The present simulations are performed on a square lattice. However, also in 
this aspect the generalization to other geometries is 
straightforward.\cite{sces05}

The outline of the paper is as follows. Section II describes the formalism.
In particular, it is demonstrated there how the classical Metropolis
algorithm can be adopted to a system with both classical and quantum
particles. In Section III, we analyze the order--disorder phase transition,
especially in the small--$U$ regime. It was suggested in Ref. \onlinecite{sces04}
that in this regime the FK model exhibits a first order phase transition. Here, 
we analyze this possibility in detail.
Section IV is devoted to the dependence of the phase--transition--temperature
on the interaction $U$. Using temperature dependence of the specific heat
and charge density wave (CDW) susceptibility we identify the critical
temperature, constructing a phase diagram in the $T-U$ plane.
In Section V, we demonstrate how the long range order vanishes when
the temperature increases. Its temperature dependence is anomalous
in a small--$U$ limit. Section VI is devoted to the ground state
and the phase diagram of the FK model with next--nearest--neighbor hopping.
Section VII presents the results for the density
of states and spectral functions of the mobile particles.
Section VIII contains our conclusions.

\section{Computational method}

In all simulations we have used the Metropolis algorithm.\cite{metropolis}
Our system contains classical (ions) as well as fermionic
(electrons) degrees of freedom. The appropriate way to treat
such a Hamiltonian is to define the grand canonical partition
function as
\begin{equation}
{\cal Z}=\sum_{\cal C} {\rm Tr}_e
e^{-\beta\left[{\cal H}({\cal C})-\mu \hat{N}\right]},
\end{equation}
where ${\cal H}({\cal C})$ is the Hamiltonian (\ref{FK})
for a fixed ionic configuration ${\cal C}$, $\sum_{\cal C}$
indicates summation over ionic configurations,
${\rm Tr}_e$ denotes the trace over fermionic degrees of
freedom, $\beta$ is the inverse temperature, and $\hat{N}$
is the operator for the total number of electrons.
For a given ionic configuration the Hamiltonian
${\cal H}({\cal C})$ can be diagonalized numerically and
the summation over fermionic degrees of freedom gives the
partition function in the following form
\begin{equation}
{\cal Z}=\sum_{\cal C} \prod_n \left\{ 1 +
e^{-\beta\left[E_n({\cal C})-\mu\right]}
\right\},
\end{equation}
where $E_n({\cal C})$ are eigenvalues of ${\cal H}({\cal C})$.
Introducing the electronic free energy
\begin{equation}
{\cal F}_e({\cal C})=-\frac{1}{\beta}\sum_n \ln
\left\{ 1 + e^{-\beta\left[E_n({\cal C})-\mu\right]}\right\},
\end{equation}
the partition function can be written in a form analogous
to that used for a spin system
\begin{equation}
{\cal Z}=\sum_{\cal C} e^{-\beta{\cal F}_e({\cal C})}.
\label{part}
\end{equation}
The above equation describes the effective model, that results
from the summation over the fermionic degrees of freedom.
As the electronic free energy ${\cal F}_e({\cal C})$ depends on
temperature, the corresponding Hamiltonian, describing only the
classical particles would include temperature--dependent interaction.
This dependence may lead to nontrivial critical exponents.
Equation (\ref{part}) indicates also, that in the Metropolis algorithm
we should use the electronic free energy in the statistical weights. To
do this, however, one has to know the value of the
chemical potential $\mu$, and apart from specific cases (e.g.,
half--filling) it has to be determined separately. Since the
calculations are carried out in the grand canonical ensemble,
the chemical potential must be kept constant during the simulation.
Determining thermodynamic quantities, the averages are calculated
with the statistical weights
\begin{equation}
w({\cal C})=\frac{1}{\cal Z}e^{-\beta{\cal F}_e({\cal C})},
\label{weight}
\end{equation}
of corresponding ionic configurations ${\cal C}$.

One of the advantages of the proposed approach is that it gives
densities of states and spectral functions for the mobile particles that
identically satisfy different sum rules. This will be discussed in Sec. VII.

The calculations
were carried out on clusters up to $10^4$ lattice sites, however,
in most cases square clusters $20\times20$ were used. Usually, the
simulations started at high temperature, with the initial state with
randomly distributed ions. Then, the temperature has been
slowly decreased. Below some temperature, ions started to
form a checkerboard patterns.

It is difficult to describe the process of pattern formation
qualitatively. In some cases it is convenient to use
$\Delta \equiv |\langle n_A \rangle -
\langle n_B \rangle|$, where $\langle n_A \rangle$ ($\langle n_B \rangle$)
is the ion concentration in sublattice A (B), as the order parameter.
However, such a description is not well suitable for cluster calculations,
since even in the case, when in the ground state ions form almost
perfect checkerboard, $\Delta$ can be close to zero. Figure 1 presents
an example of such a ionic configuration.
\begin{figure}
\includegraphics[width=3.7cm,clip=true]{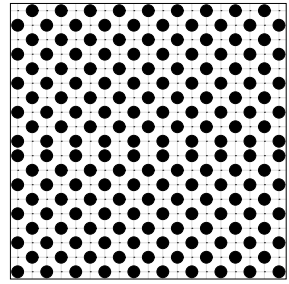}
\caption{Example of almost perfect checkerboard ionic configuration,
for which the long range order parameter $\Delta$ is close to zero.
Filled circles represent sites occupied by the massive particles.}
\label{fig1}
\end{figure}
A small staggered
field can be introduced in order to prevent an occurrence of such effects.
However, this ``phase smoothing out'' may restrict also some other
configurations.

Instead, we use the density--density correlation function for
the ions to describe the ordered state. It has the advantage that
it is capable to describe long range as well as short range correlations.
We define the correlation function in the following way:
\begin{equation}
g_n=\frac{1}{4N}\sum_{i=1}^{N} \sum_{\tau_1,\tau_2=\pm n}
w\left({\bm r}_i\right)w\left({\bm r}_i+\tau_1{\bm\hat{x}}
+\tau_2{\bm\hat{y}}\right),
\label{cor_fun}
\end{equation}
where $w\left({\bm r}_i\right)\equiv w_i$, ${\bm\hat{x}}$, and
${\bm\hat{y}}$ denotes unitary vectors
along the $x$ and $y$ directions, respectively (the lattice
constant $a=1$ has been assumed). Note, that $g_n$ describes
correlations along the lattice axes.

The correlation function $g_n$ describes ionic configuration only.
However, the distribution of the light particles correlates with
the ionic distribution due to the direct interaction between these
two types of particles. The larger value of $U$ the stronger correlations
take place. Figure 2 demonstrates these correlations for weak (upper
configurations) and strong (lower configurations) interaction.
\begin{figure}
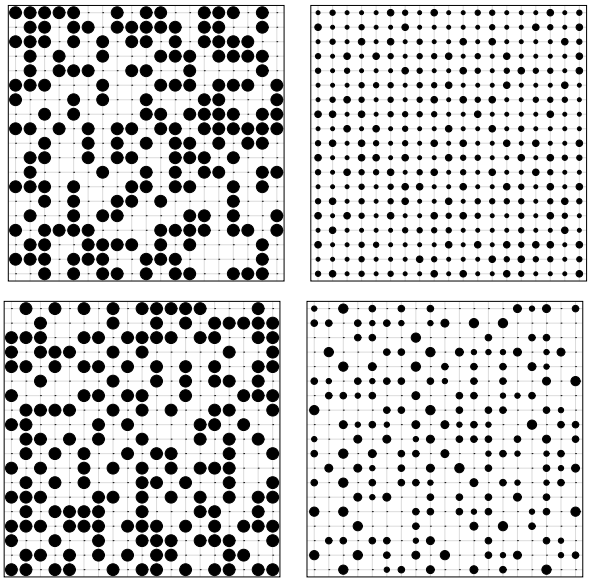

\centerline{
\includegraphics[width=3.7cm,clip=true]{fig2a.eps}
\ \
\includegraphics[width=3.7cm,clip=true]{fig2b.eps}}
\vspace*{2mm}
\centerline{
\includegraphics[width=3.7cm,clip=true]{fig2c.eps}
\ \
\includegraphics[width=3.7cm,clip=true]{fig2d.eps}
}
\caption{Examples of ionic (left) and electronic (right) configurations
for weak ($U/t=0.5,\ k_{\textrm B}T/t=0.1$, upper row) and strong
($U/t=5,\ k_{\textrm B}T/t=0.3$, lower row) interaction. Diameters
of the circles in the right configurations are proportional to the
concentration of electrons.}
\end{figure}

If there is a phase transition, some thermodynamical quantities have to
diverge at the critical temperature. Here, we used the CDW susceptibility
$\chi$  and the specific heat $C_V$ to determine this point.
The CDW susceptibility is related through the fluctuation--dissipation theorem
to the variance of the density--density correlation function. This is a form
especially convenient for the MC calculations
\begin{equation}
\chi=
\frac{1}{k_{\rm B}T}\left(\langle g_1^2 \rangle - \langle g_1 \rangle^2 \right),
\label{chi}
\end{equation}
where we used $\langle \ldots \rangle$ to indicate the average over generated
ionic configurations. It is a little bit more complicated to determine the
specific heat. Within the standard MC approach the specific heat is calculated
from the fluctuations of the energy, similarly to Eq. (\ref{chi}). In our case,
however, according to Eq. (\ref{weight}) we use the electronic free energy in
the Metropolis algorithm and the internal energy is not directly
available from the simulations. On the other hand, the Fermi energy is
much larger than the order--disorder transition temperature. Therefore,
estimating the specific heat we replace the trace over the fermionic degrees
of freedom by a ground state expectation value $E$,\cite{trace} which is
a temperature--independent quantity. Then, the specific heat can be determined
from the fluctuation--dissipation theorem
\begin{equation}
C_V=
\frac{1}{N}\frac{1}{k_{\rm B}T^2}\left(\langle E^2 \rangle - \langle E
\rangle^2 \right).
\label{cv1}
\end{equation}

\begin{figure}
\includegraphics[width=8.5cm,clip=true]{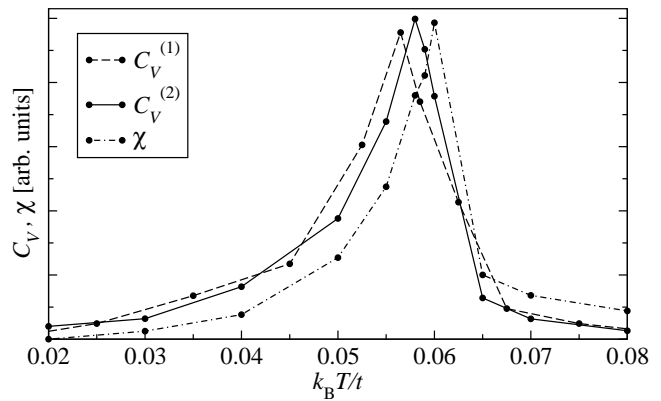}
\caption{Specific heat determined from Eq. (\ref{cv1}) ($C_V^{(1)}$, dashed line)
and from Eq. (\ref{cv2}) ($C_V^{(2)}$, solid line) for $U/t=1$. For comparison,
CDW susceptibility $\chi$ is also presented (dashed--dotted line).}
\label{test_cv}
\end{figure}

In order to confirm the validity of this approximation we compare
the specific heat determined numerically from the relation
\begin{equation}
C_V=\frac{dU}{dT}
\label{cv2}
\end{equation}
and that from Eq. (\ref{cv1}).
Since the simulations were carried out only for a given set of temperatures,
a finite difference approximation had to be used for the derivative in Eq. (\ref{cv2}).
Nevertheless, within the temperature regime of interest the difference
is below a few percent. Figure \ref{test_cv} shows a comparison of the specific
heat calculated from Eq. (\ref{cv1}) and from Eq. (\ref{cv2}). Additionally,
the CDW susceptibility is plotted there in order to indicate the critical temperature.
A comparison of the critical temperatures determined from the specific heat and
from the CDW susceptibility, presented in Figs. (\ref{fig7}--\ref{fig9}),
confirms the validity of the proposed approach as well. Note, that these quantities
are calculated in completely different ways: the specific heat is determined from
the eigenvalues of the fermionic Hamiltonian, whereas
the CDW susceptibility from the correlation function of the classical particles.

In each simulation
the ensemble averages of the thermodynamic quantities of interest
are calculated after the system equilibrated. The thermalization
period varies with temperature. In particular, it is
very long at low temperatures or in the vicinity of critical
points (the critical slowdown). In order to determine whether the system
has reached equilibrium,
we usually start the simulation with two replicas of the system:
one starting from a fully ordered state, the other from a random
ion distribution. Then, the energies of the systems
evolve crossing each other after some number of MC steps.
This point is considered as the beginning of equilibrium state and the
averages of the thermodynamic quantities are calculated over the
remaining MC steps. This procedure is especially useful
close to the transition temperature.

The width of the peak in the specific heat and in the CDW susceptibility
decreases with the increase of the size of the system and a finite--size 
scaling should be carried out in order to determine precisely the transition 
temperature. This can be done, for instance, using the standard Binder cumulant
method.\cite{binder1} However, taking into account the huge amount of time 
needed to run the finite--size scaling over the whole parameter space, we have
decided to omit to do it and most of the presented results have been obtained 
for 20 $\times$ 20 cluster. Such a cluster is sufficiently large to produce
results accurate enough to describe the properties of the model under 
investigation and, on the other hand, the time of a single simulation run 
is short enough to allow determination of the full phase diagram. Moreover,
some comparison of results obtained for 20 $\times$ 20 and 40 $\times$ 40 
clusters is presented in Section VII.

\section{Nature of the phase transition}

For large $U$ there
is an Ising--like phase transition at the critical temperature
$T_c \propto U^{-1}$. In the small--$U$ regime the critical
temperature is bounded from below by $U^2/|\ln U|$.\cite{Ken-Lieb1,Ken-Lieb2}
In fact, it was shown that in large dimensions
$T_c \propto U^2|\ln U|$ for $U\rightarrow 0$.\cite{infD2}

It is known that for large $U$ the Falicov--Kimball model belongs
to the same universality class as the Ising model and the order
parameter is described by Curie--Weiss law
$\Delta=\tanh\left(\Delta/\Theta\right)$, where $\Theta=T/T_c(U)$.
Thus, the phase transition is of second order in this regime.
On the other hand, it was shown in Refs. \onlinecite{infD2}
and \onlinecite{CFJ},
that in infinite dimensions in the small--$U$ limit the order
parameter has a strange non--BCS--like temperature dependence.
Therefore, it is important to determine precisely the nature of
the phase transition in this regime. We have used a method
proposed by M. Challa {\em et al.}\cite{binder} to
distinguish between first and second order phase transitions.
Systems undergoing first order phase transitions are accompanied
by free energy barriers which separate the free energy minima
characterizing the coexisting phases. It results in
discontinuities in the first derivatives of the free energy, e.g.,
the internal energy. This, in turn, leads to a $\delta$--type
singularity in the specific heat at the transition.

Within the framework of Metropolis MC approach the
energy $E$ fluctuates with the probability distribution
$P(E)$ usually given by a Gaussian. Its width is proportional
to the specific heat. However, if the system is close to the
first order transition, the probability distribution $P(E)$
is a superposition of two Gaussians centered at different
energies, $E_+$ and $E_-$. Here, $E_+$ and $E_-$ are the
energies in the high-- and low--temperature phases,
respectively. The Gaussians are weighted by the Boltzmann
factors of $E_+$ and $E_-$, and thus this splitting occurs
in a vicinity of the transition temperature only. At higher
(lower) temperatures $P(E)$ forms a single Gaussian centered
at $E_+$ ($E_-$).

\begin{figure}
\includegraphics[width=8.5cm,clip=true]{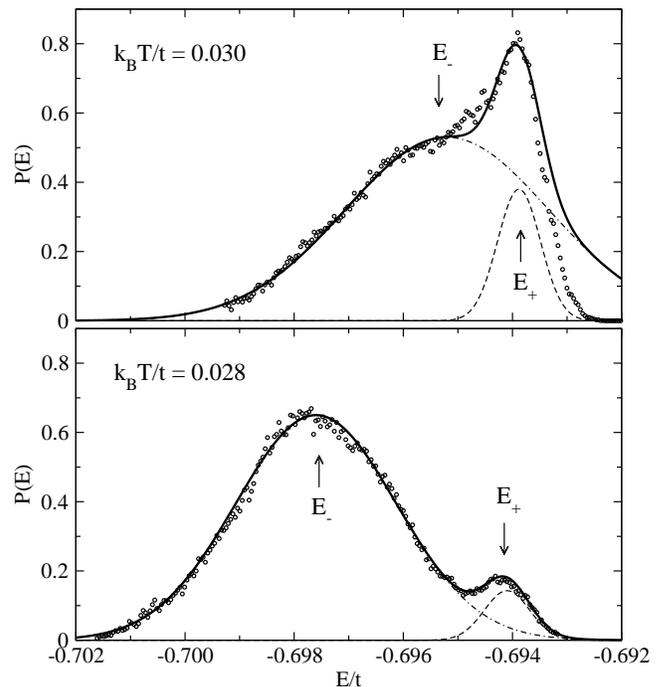}
\caption{Probability distribution of the energy at
temperatures close to the critical temperature for $U/t=0.5$.
The thick solid lines represent a superposition of two Gaussians
that fit the simulation results, whereas the dashed and
dashed--dotted lines show the component
Gaussians. The arrows indicate positions of the centers of
the Gaussians representing low--temperature ($E_-$), and
high--temperature ($E_+$) phases, respectively.}
\label{fig2}
\end{figure}

Figure \ref{fig2} shows the energy distribution
for temperatures slightly below and above the transition temperature for $U/t=0.5$. 
Since the data has a double--peak structure, the phase
transition is of the first order. One can see the transfer
of the weights from the low--temperature Gaussian peak
(the lower panel in Fig. \ref{fig2}) to the high--temperature
one (the upper panel), as the temperature increases, passing
through the critical value. The energy distribution at the critical
point, where the heights of both the peaks are comparable, has been presented 
in Ref. \onlinecite{sces04}.

\begin{figure}
\includegraphics[width=8.5cm,clip=true]{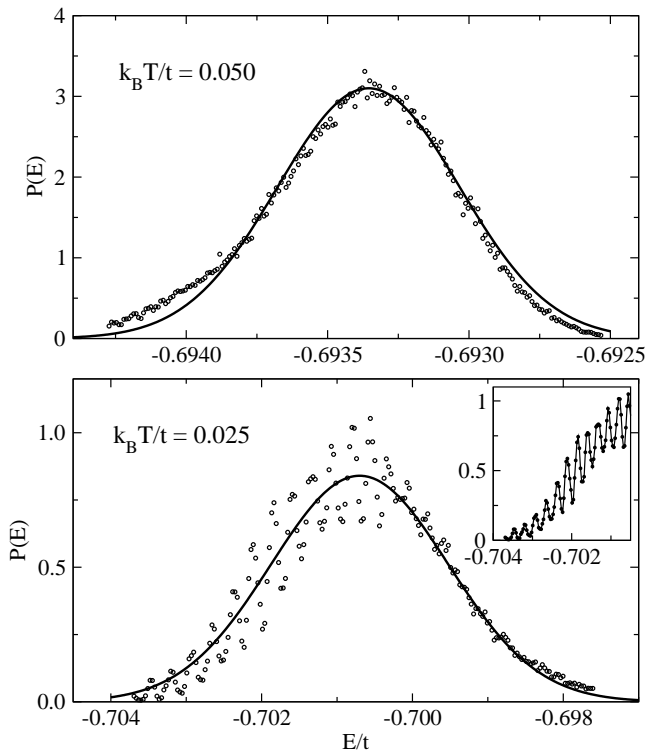}
\caption{Probability distribution of the energy at
temperatures much higher and much lower than the transition
temperature for $U/t=0.5$. Inset in the lower panel shows the
low--energy part of the distribution with lines connecting
the points.}
\label{fig3}
\end{figure}

At temperatures much lower and much higher than the transition
temperature, the energy distribution can be well fitted by a
single Gaussian. Such situations are presented in Fig. \ref{fig3}.

The energy distribution at low temperature (lower panel in Fig.
\ref{fig3}) consists of a large number of peaks, clearly visible
in the inset. They are connected with the discrete spectrum
of the Hamiltonian for a finite system. Namely, at such a low
temperature, there are only a few dislocations in the checkerboard
ionic configuration. Due the low concentration of the dislocations,
they are almost independent and each of them changes the
energy by an approximately the same amount. In this way, two successive
peaks correspond to configurations with the numbers of dislocations
that differ by one. As the number of dislocations increases, they
start to ``feel'' each other, and the effective interactions smear out
this energy ladder. This is why the peaks are visible only in the
low--energy part of the distribution. Of course, when the size of the
lattice increases, the energy spectrum becomes quasicontinuous and
the oscillations disappear.

\begin{figure}[h]
\includegraphics[width=8.5cm,clip=true]{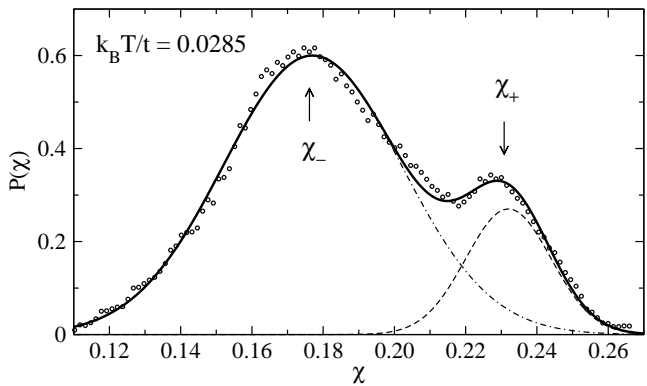}
\caption{Probability distribution of the CDW susceptibility $\chi$ close to
the phase transition for $U/t=0.5$. The meaning of the lines is analogous to
that of Fig. \ref{fig2}.}
\label{fig_cdw_hist}
\end{figure}

The coexistence of low-- and high--temperature phases at the phase transition
can also be observed in the CDW susceptibility. Fig. \ref{fig_cdw_hist}
shows the two--peak structure of the probability distribution of $\chi$.
It should be noted that the distribution presented in Fig. \ref{fig_cdw_hist}
describes ionic configurations, whereas the one presented in Fig. \ref{fig2}
is obtained from the eigenvalues of the fermionic Hamiltonian. The similarity
between these two distributions speaks strongly in favor of the validity
of the proposed MC approach, confirming the presence of first order phase
transition in the small--$U$ regime.

\begin{figure}[h]
\includegraphics[width=8.5cm,clip=true]{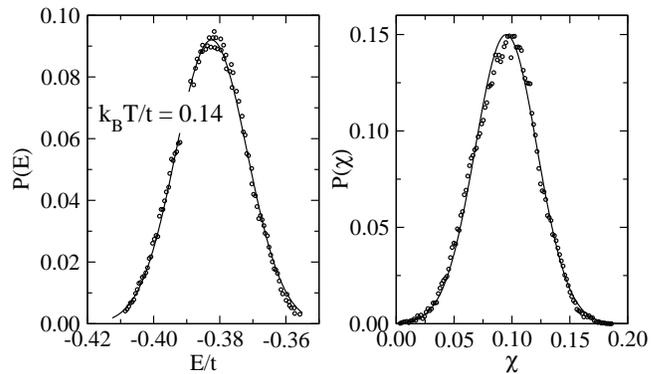}
\caption{Probability distribution of the energy (left panel) and of the CDW 
susceptibility (right panel) at the phase transition for $U/t=3$.}
\label{fig4}
\end{figure}

The magnitude of splitting of the energy distribution close to $T_c$ decreases
with the increase of the interaction strength $U$ and above a critical
value $U^*$ disappears. The same holds true for the distribution of the CDW
susceptibility. Figure \ref{fig4} presents the energy and CDW susceptibility 
distribution at the phase transition for $U/t=3$. One can notice an excellent 
agreement with the theoretical Gaussian curve, indicating an absence of any 
phase coexistence and the second order character of the phase transition.

We have estimated the critical value $U^* \approx t$, however,
extensive numerical studies are necessary in order to
determine $U^*$ precisely.

\subsection{Phase diagram}

\begin{figure}[h]
\includegraphics[width=8.5cm,clip=true]{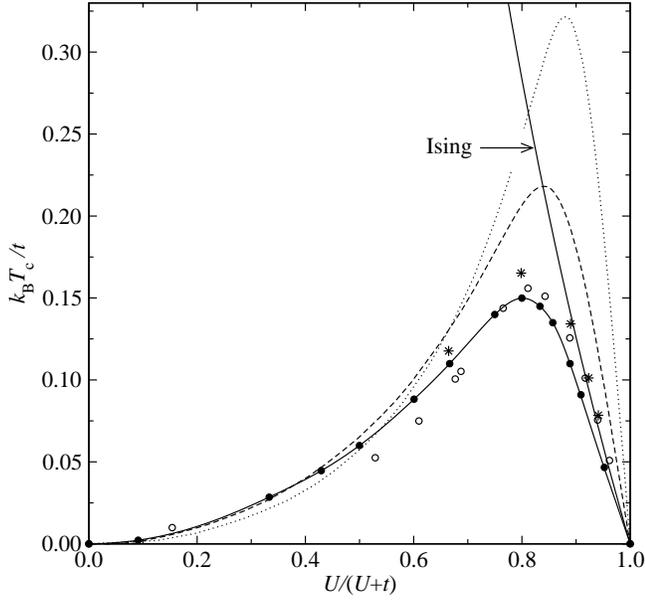}
\caption{Transition temperature for the $d=2$ half--filled Falicov--Kimball
model as a function of the interaction strength (solid line with circles).
The dotted and dashed lines present
for comparison results taken from Ref. \onlinecite{CFJ}.
These results were obtained in $d=\infty$ limit
for the Bethe lattice (dotted line) and the hypercubic lattice (dashed
line). Open circles represent numerical results taken from Ref.
\onlinecite{Raedt},
estimated as the temperature at which a gap in the density of states closes.
Stars represent results obtained in Ref. \onlinecite{DCA} from DCA with a QMC
for a 36-cite cluster. 
The solid line connecting filled circles is a guide for the eyes only.
}
\label{fig5}
\end{figure}

The position of peaks in the specific heat and the CDW susceptibility
has been used to determine the critical temperature. Fig.
\ref{test_cv} shows the typical temperature dependence of the specific heat. 
The phase diagram has been constructed using results obtained for 
$20\times 20$ system. Figure \ref{fig5} presents
the transition temperature as a function of the interaction strength.
The horizontal axis is plotted as $U/(U+t)$, such a scaling allows one
to present both the weak and strong coupling results in the same graph.
There are also shown results obtained by means of other methods: 
DMFT for $d=\infty$ case, taken from Ref. \onlinecite{CFJ}, MC taken from
Ref. \onlinecite{Raedt}, and DCA taken from Ref. \onlinecite{DCA}. In the large--$U$ 
limit the critical temperature for the Ising model is also presented for
comparison.


\section{Order parameter}

\begin{figure}[h]
\includegraphics[width=8.5cm,clip=true]{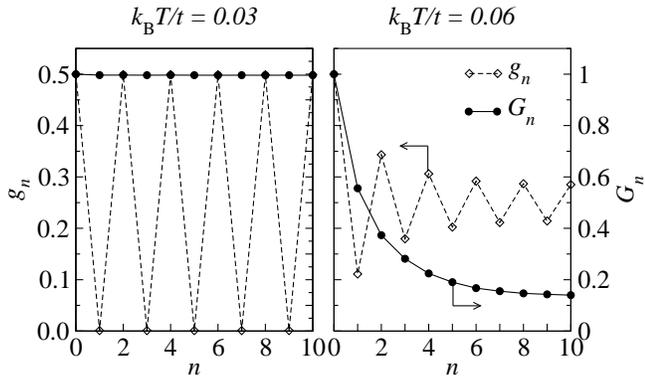}
\caption{Correlation functions $g_n$ (dashed lines) and $G_n$
(solid lines) for $U/t=20$. Left
panel shows the case of a fully ordered (checkerboard) state
and the right one corresponds to the vicinity of the phase transition.}
\label{fig6}
\end{figure}

In order to quantitatively describe the ionic configuration, we have
investigated the density--density correlation function defined in Eq.
(\ref{cor_fun}). This function is capable of descring short range
as well as long range order. In a fully ordered (checkerboard) state
it oscillates with a period equal to two lattice constants (see Fig.
\ref{fig6}).

\begin{figure}[h]
\includegraphics[width=8.5cm,clip=true]{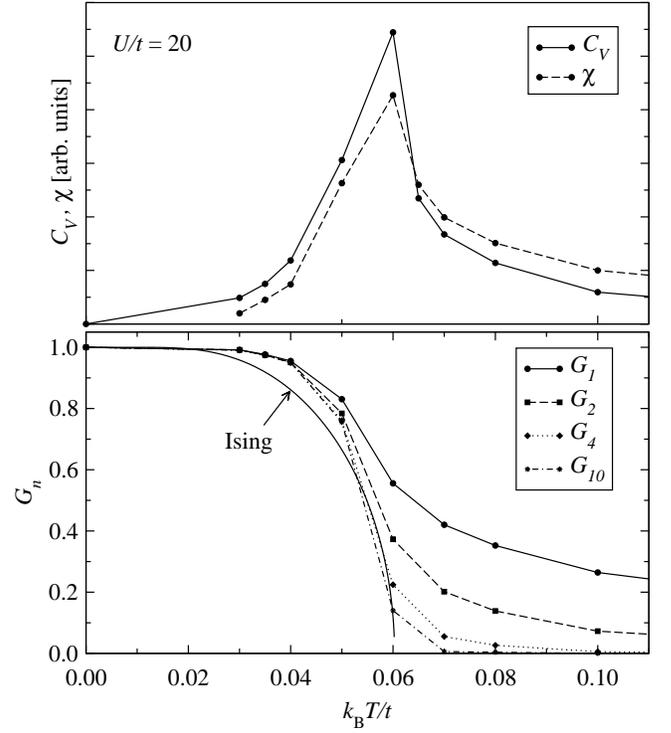}
\caption{Specific heat $C_V$ and CDW susceptibility $\chi$ (upper panel)
and correlation function $G_n$ (lower panel) as a function of temperature
for $U/t=20$. Various lines in the lower panel correspond to correlation
functions $G_n$ calculated for various distances $n$. For comparison,
there is also a line representing temperature dependence of the magnetization
in the Ising model}
\label{fig7}
\end{figure}

\begin{figure}[!]
\includegraphics[width=7.7cm,clip=true]{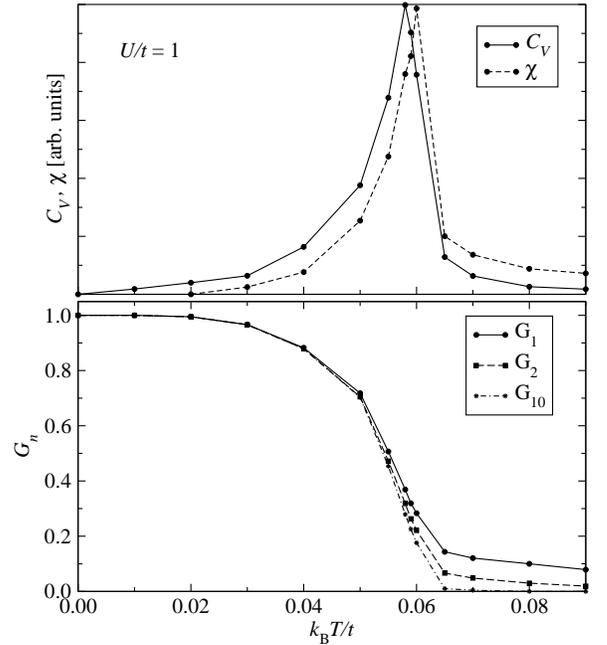}
\caption{The same as in Fig. \ref{fig7} but for $U/t=1$.}
\label{fig8}
\end{figure}

\begin{figure}[!]
\includegraphics[width=7.7cm,clip=true]{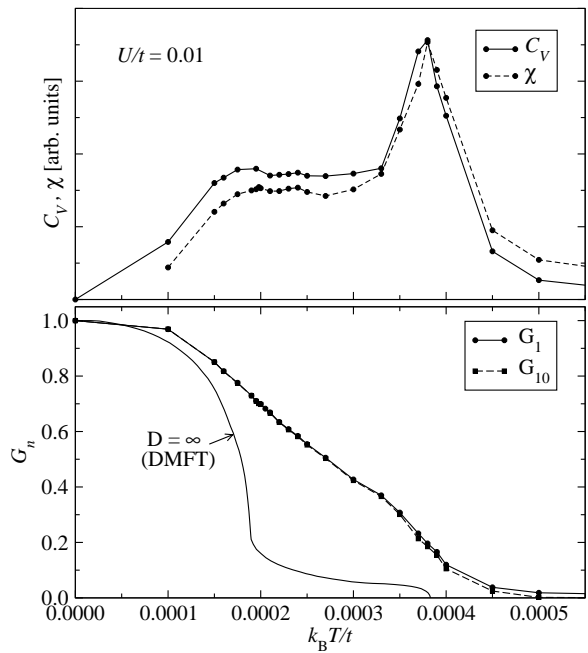}
\caption{The same as in Fig. \ref{fig7} but for $U/t=0.01$.
Additionally, there is also a line representing the order parameter
determined from DMFT solution of the FK model in infinite dimensions
in the limit $U\rightarrow 0$.
This solution is taken from Ref. \onlinecite{vanDongen}.
}
\label{fig9}
\end{figure}

It is convenient to define a renormalized correlation function
$G_n=(-1)^n4\left(g_n-\rho_i^2\right)$, $\rho_i\equiv N_i/N$ is the
concentration of ions, where $N$ and $N_i$ are numbers of lattice
sites and ions, respectively.
Such a function is equal to 1 for the checkerboard
state and close to 0 for randomly distributed ions, independently of the
distance $n$ (for $n>0$).
Apart from these limiting cases, this
function decreases monotonically with increasing distance.
Figures \ref{fig7}--\ref{fig9} present the temperature
dependence of $G_n$ for a wide range of the interaction $U$.
These curves are presented together with the temperature dependencies
of the specific heat and the CDW susceptibility, calculated for the same
values of $U$. This allows one to find the exact values of the critical
temperatures.

For strong to intermediate values of the Coulomb interaction there
is a distinct peak in the specific heat, indicating the phase
transition from the ordered state to the disordered one. The
corresponding vanishing of the correlation function $G_n$
resembles a typical behavior of an order parameter close to
the second order phase transition (see Figs. \ref{fig7} and \ref{fig8}).
Note, that the long range correlations disappear more rapidly than
that for shorter distances. This can be, however, attributed to the
finite size of the system.

On the other hand, in the weak interaction
limit the correlation functions
vanishes almost linearly and the slope begins already at temperature
approximately $0.25\:T_c$.
An unusual behavior in this regime is seen also in
the specific heat: there is hump in both $C_V(T)$ and $\chi(T)$
close to the temperature at which the slope in $G_n(T)$ occurs.
Fig.~\ref{fig9} illustrates such a behavior for $U/t=0.01$. For
weaker interaction the values of the specific heat and the CDW
susceptibility are different, however, the shape of their temperature
dependence, as well as the temperature dependence of the correlation
function, remain almost unchanged.

Such an anomalous temperature evolution of $G_n(T)$ may be
explained as a result of finite size of the system. For small
values of $U$ there exist solutions with periodic ionic configurations
possessing very large period.
Such configurations are excluded from our calculations due
to limited size of the clusters the simulations were carried out on.

On the other hand, it was recently shown that similar behavior
of the order parameter occurs in the infinite--dimensional limit.
\cite{vanDongen,CFJ}
Our results may suggest that such a behavior is a generic property
of the weak--coupling Falicov--Kimball model.

\section{Next--nearest--neighbor hopping}

\subsection{Ground state}

In this section we generalize the FK Hamiltonian by taking into account
the hopping to next nearest neighbors (NNN) with the hopping integral $t'$ in addition
to the nearest--neighbor (NN) hopping. In particular, the question concerning the nature 
of the ordered state is addressed. In a more general case of various concentration
of both type of particles this problem was analyzed in Ref. \onlinecite{Wojtkiewicz}.

\begin{figure}[!]
\includegraphics[width=6.5cm,clip=true]{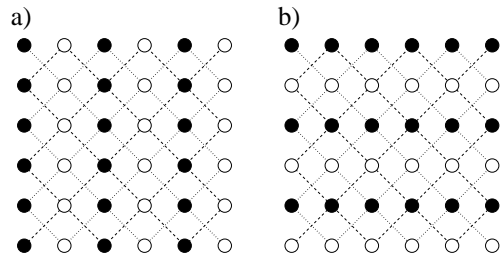}
\caption{Possible ground--state ionic configurations in $t'\gg t$ limit. The dashed and
dotted lines indicate (independent in $t\rightarrow 0$ limit) sublattices.}
\label{fig_nnn_conf}
\end{figure}

In the limiting case of $t'\ll t$ one may expect that the
checkerboard state is still the actual ground state of the FK model. In the opposite
limit, when the hopping to the nearest sites can be neglected, the square lattice
can be divided into two interpenetrating square sublattices with the lattice constant
$\sqrt{2}$ times larger than the original one and with the axes rotated by $45^\circ$.
Since the electrons do not hop between the sublattices, the system breaks up into two
uncoupled, interpenetrating square lattices composed of NNN bonds. Each of these
lattices is independently described by the FK Hamiltonian. Since both the
sublattices are bipartite and both the subsystems
are half--filled, the ions will arrange into checkerboard patterns in the ground state.
Depending on the relative phase between the orderings in the sublattices, the resulting
ground state of the whole system will have the form of vertical or horizontal stripes.
Fig. \ref{fig_nnn_conf} demonstrates these possibilities.
In the large--$U$ limit the threshold value of the ratio $t'/t$, that separates
the ground states with ions forming the checkerboard and stripe patterns, can be
determined using the mapping of the FK model onto the Ising one. In the antiferromagnetic
Ising model with NN and NNN coupling the ground state is the simple antiferromagnet for
$J'/J < 0.5$, where $J$ and $J'$ are the NN and NNN interactions, respectively.
For $J'/J>0.5$ the system minimizes the energy by ordering in alternate ferromagnetic
rows of opposite spins.\cite{landau, landau_binder, moran} Such spin configurations
(``superantiferromagnetic'') correspond to the ionic configurations of the FK model
presented in Fig. \ref{fig_nnn_conf}. Since the ratio $J'/J$ of Ising NN and NNN interactions
is equal to $(t'/t)^2$ in the corresponding FK model, the threshold value of the NNN
hopping is given by $t'/t\simeq0.71$. 
%
In Ref. \onlinecite{Wojtkiewicz} it has been shown that this result holds true up to fourth 
order in $t/U$.
For weaker interaction this threshold ratio can be determined from a comparison
of the ground state energy of both the checkerboard and stripe configurations.
Our simulations indicate that the critical value of $t'/t$ changes slightly with $U$, 
however, it always lies between 0.71 and 0.8.

\subsection{Specific heat}

\begin{figure}[!]
\includegraphics[width=7.7cm,clip=true]{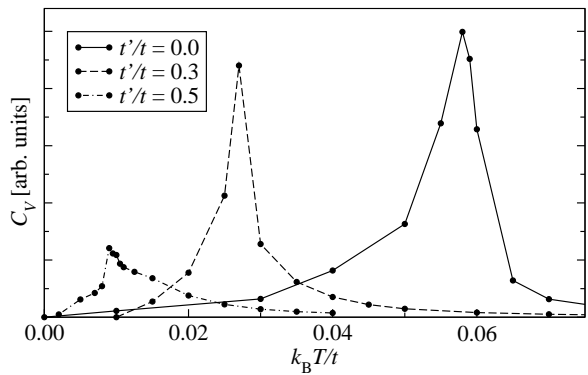}
\caption{Specific heat for $U/t=1$ and $t'/t = 0,\ 0.3$ and 0.5.}
\label{fig_nnn_spec_heat}
\end{figure}

\begin{figure}[!]
\includegraphics[width=7.7cm,clip=true]{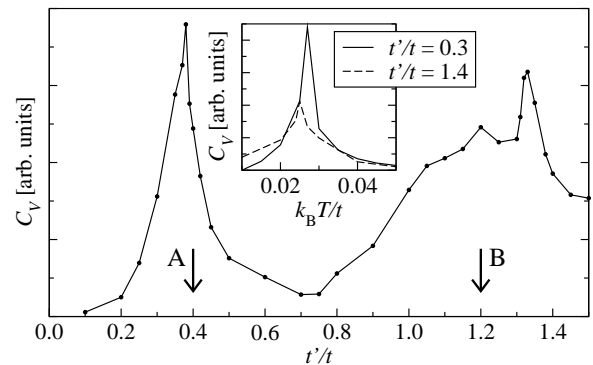}
\caption{Specific heat for $U/t=1$ as a function of $t'/t$ at $k_{\rm B}T/t=0.02$.
The arrows indicate values of $t'$, for which $k_{\rm B}T/t=0.02$ is the critical
temperature of the phase transition from the checkerboard state (A) and from the
stripe phase (B).
The inset shows temperature dependence of the specific heat for $t'/t=0.3$ and $t'/t=1.4$.}
\label{fig_nnn_spec_heat1}
\end{figure}

\begin{figure}[!]
\includegraphics[width=7.7cm,clip=true]{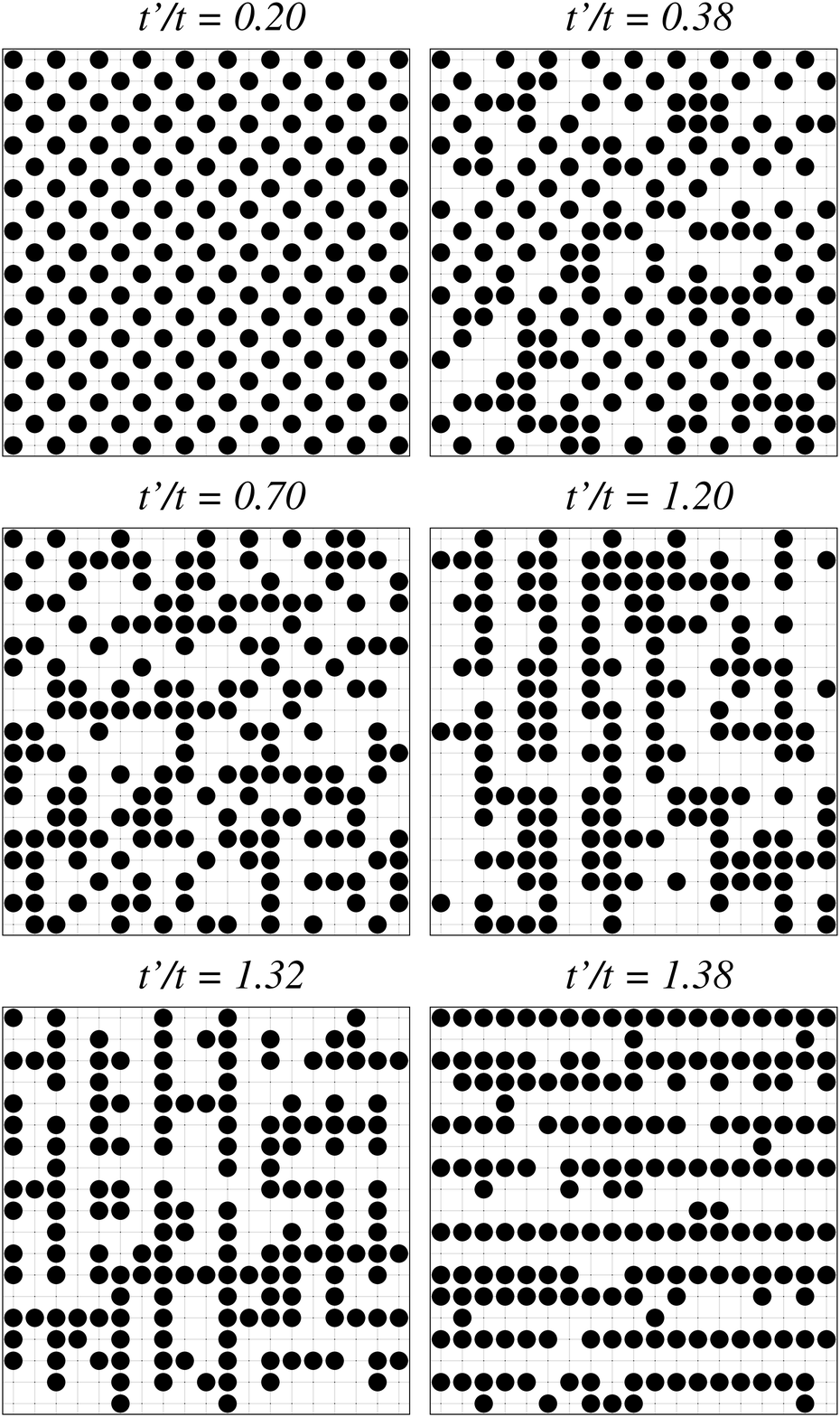}
\caption{Snapshots of ionic configurations for various NNN hopping amplitudes. The simulations
were performed for $U/t=1$ and $k_{\rm B}T/t=0.02$.}
\label{configs}
\end{figure}

Since the NNN hopping term introduces frustration, it reduces the temperature of the
transition from the checkerboard to the disordered state (for $t'/t < 0.71$). In the strong
interaction limit the dependence of the critical temperature on $t'/t$ can be determined
from results obtained for the corresponding Ising model with NNN interaction. For finite
$U/t$ the critical temperature can be identified from the position of the peak in the
temperature dependence of the specific heat. Figure \ref{fig_nnn_spec_heat} presents results
for $U/t=1$.
The critical temperature decreases with increasing $t'$, going to zero for $t'/t\simeq 0.71$.
At this point there is no long range order at any finite temperature. Then, the critical temperature
increases with further increasing $t'$. In this regime the peak in the specific heat
indicates the transition from the stripe configuration to the disordered state.
For $t'/t\gg 1$ the nearest--neighbor hopping can be
neglected and one ends up with the FK model on a square lattice with NN hopping integral $t'$.
Therefore, according to the phase diagram presented in Fig. \ref{fig5}, in the limit
$t'/t \rightarrow \infty$ (for a given $U$) the critical temperature goes to zero.
Figure \ref{fig_nnn_spec_heat1} shows how the specific heat depends on $t'/t$ at
temperature $k_{\rm B}T/t=0.02$.
This is a critical temperature for two different values of $t'$:
one for $t'/t=0.4$, when in the low--temperature phase the ions form the checkerboard
pattern, and the other for $t'/t=1.2$, when stripes minimize the free energy at low
temperature. As a result, the specific heat plotted as a function of $t'/t$ has two
maxima, as can be seen in Fig. \ref{fig_nnn_spec_heat1}. The left peak corresponds to the 
transition from the checkerboard state, whereas the right one
to the transition from the stripe state. The widths of the peaks are connected with the
effective sizes of the lattices. For large NNN hopping the lattice breaks up into two
sublattices,
which becomes decoupled in the limit of $t'/t \rightarrow \infty$ and the checkerboard pattern
is formed in each of them independently. Therefore, the simulations are effectively
carried out for two replicas of a system of halved size. As a result, the system is further
away from the thermodynamic limit than in the case of small $t'$ and the maximum of the 
specific heat is less sharp (see the inset in Fig. \ref{fig_nnn_spec_heat1}). The features
seen at the top of the second maximum are connected with a rearrangement of the ions from 
two--dimensional structures to one--dimensional ones, as can be seen in the lowest row of 
Fig. \ref{configs}. Figure \ref{configs} illustrates how the ionic configuration evolves from
the checkerboard to stripe pattern with increasing amplitude of the NNN hopping. Due to the
degeneracy of the ground state for $t'/t >0.71$, showed in Figure \ref{fig_nnn_conf}, the
evolution goes with equal probability through a state with vertical or horizontal stripes.
Finally, for any finite temperature the system ends up in a disordered state. 

\subsection{Phase diagram}

The determination of the full phase diagram of the FK model with NNN hoppings is a very
CPU time consuming task due to the large number of free model parameters. Therefore, we have
determined only a few points in order to qualitatively describe the dependence of the
critical temperature on the ratio $t'/t$. \begin{figure}[!]
\includegraphics[width=7.7cm,clip=true]{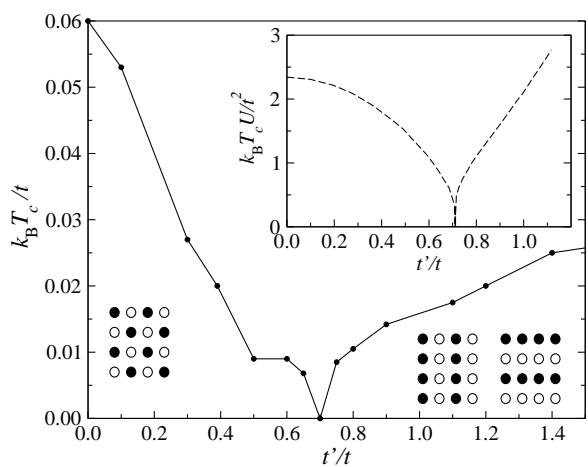}
\caption{Critical temperature as a function of $t'/t$ for $U/t=1$. The line connecting points
is a guide for the eyes. The configurations indicate the ground states in given
$t'/t$ regimes.
The inset shows the critical temperature in $U/t\rightarrow\infty$ limit.}
\label{phase_diag}
\end{figure}
Figure \ref{phase_diag} shows the results for $U/t=1$.
This figure does not show the decrease of the critical temperature, when
$t'$ becomes very large. This is because we show the temperature in units of the NN hopping
integral $t$. In the case of small NNN hopping, $t$ is directly connected to the band width.
However, when $t'$ is very large, the band width is connected to $t'$ rather then to $t$
and $k_{\rm B}T_c/t'$, instead of $k_{\rm B}T_c/t$, would go to zero when $t'$ goes to infinity.
The inset demonstrates the corresponding phase diagram in $U/t\rightarrow\infty$ limit
obtained through mapping onto the NNN Ising model. Since the critical temperature goes to zero
when $U$ goes to infinity (for a given bandwidth), the temperature has been plotted as
$k_{\rm B}TU/t^2$.

\section{Density of states}

\begin{figure*}
\centerline{
\includegraphics[width=17.5cm,clip=true]{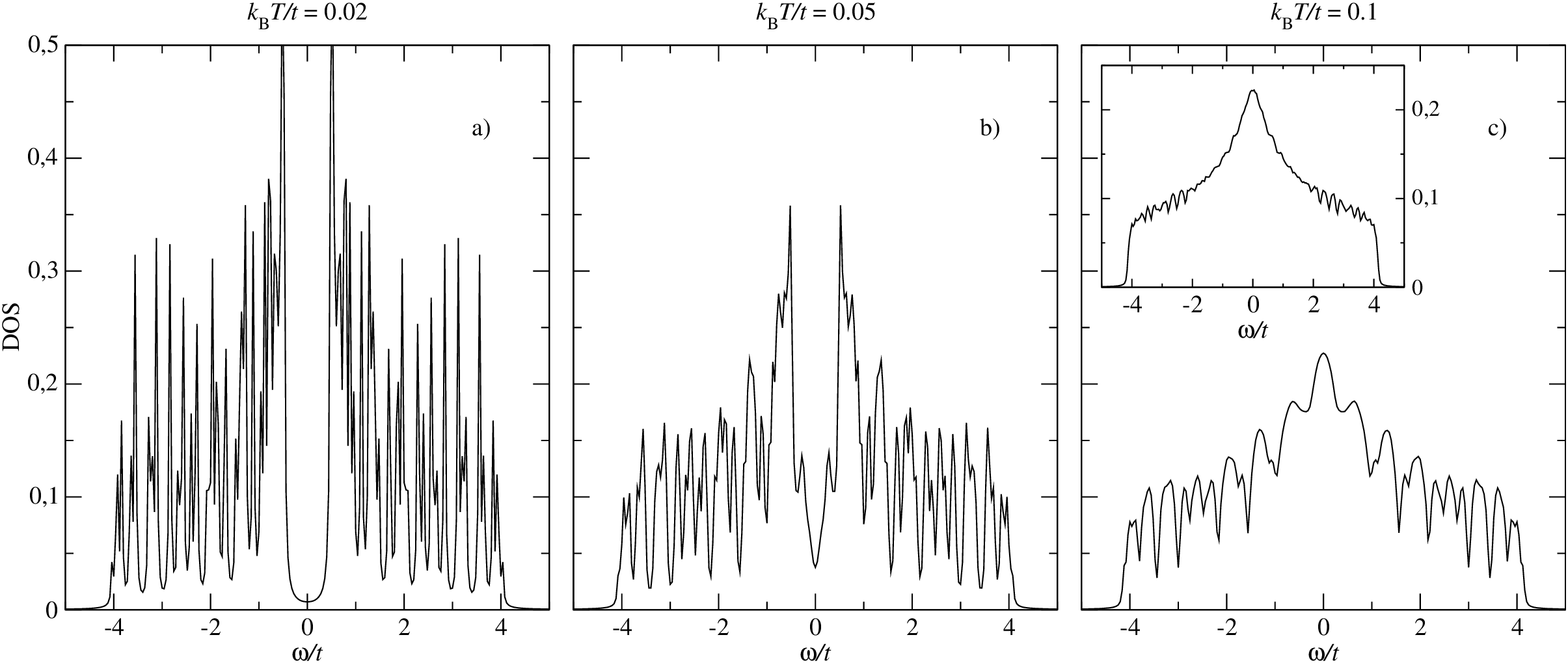}
}
\caption{Density of states for $U/t=1$ at various temperatures.
For temperatures $T$ higher then $k_{\textrm B}T/t=0.1$ the DOS is
almost the same as in panel c). The inset in panel c) shows the corresponding 
DOS for 40 $\times$ 40 cluster.
\label{DOS_smallU}}
\vspace*{4mm}
\centerline{
\includegraphics[width=17.5cm,clip=true]{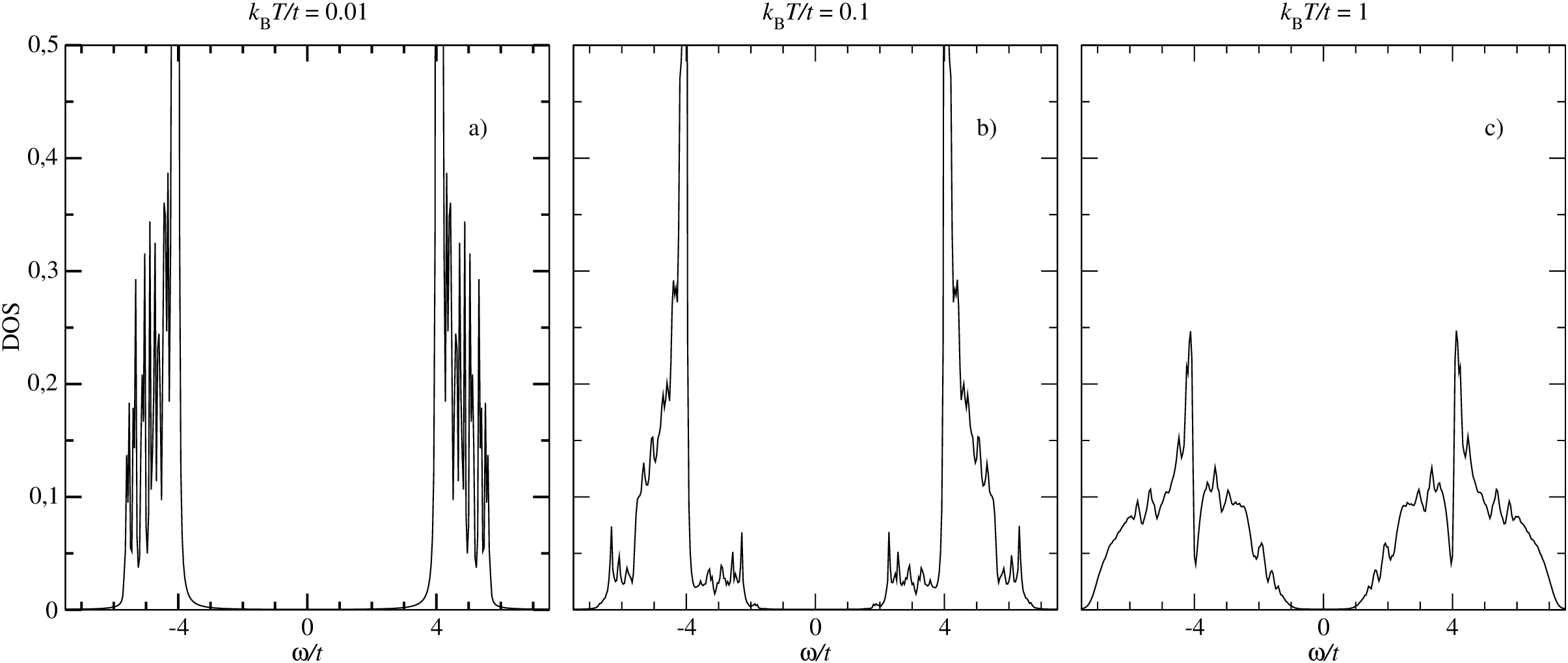}
}
\caption{Density of states for $U/t=8$ at various temperatures.
For temperatures $T$ higher then $k_{\textrm B}T/t=1$ the DOS is
almost the same as in panel c).
\label{DOS_largeU}}
\end{figure*}

Since the electron distribution correlates with the ionic configuration,
the formation of the checkerboard pattern results in a modification of
the electronic density of states (DOS). The DOS at a given temperature can
be obtained
by averaging densities determined in each MC sweep. For weak electron--ion
interaction at high temperature the electrons on average are almost
unaffected by the ionic configurations and its DOS resembles that for
free electrons on square lattice, except for the lack of the van Hove
singularity (see Fig. \ref{DOS_smallU}c).
As the
temperature is lowered the ions form the checkerboard pattern and
the electronic concentration decreases in sites that are occupied by
the ions. As a result of this CDW order a gap opens in the DOS at 
the Fermi level (see Figs.
\ref{DOS_smallU}a and \ref{DOS_smallU}b). The corresponding metal--insulator
transition remains in accordance with the Mott picture.\cite{Mott}
The situation is different in the strong interaction regime. In this case,
the electron--ion interaction almost completely forbids electrons
from occupying sites that are already occupied by the ions. Therefore, even
at high temperature
the DOS averaged over the MC sweeps differs from the free electron DOS.
At high enough temperature the Metropolis algorithm accepts every ionic
configurations with probability almost equal to one. In other words,
at high temperatures the ionic configurations are not affected by the
electron--ion interaction and the ions are distributed randomly. In this 
regime, the FK model reduces itself to a model of
free electrons in a random potential of disordered binary alloy. For
large enough $U$ there is a gap in the DOS, that does
not change further with the increasing temperature (see Fig.
\ref{DOS_largeU}c). When the temperature is lowered, this
gap transforms itself into the CDW gap (Figs. \ref{DOS_largeU}a and
\ref{DOS_largeU}b).

There is a significant difference in the temperature evolution of
the DOS between the FK model and a CDW system with selfconsistently determined
gap.\cite{gruner} In the FK model the distance between the edges of the gap 
is constant when the temperature increases, but its depth as well as the high 
of the peaks decreases. It will be clearly visible in the proceeding section, 
where the spectral functions are presented. On the contrary, the width of the
selfconsistently determined CDW gap decreases with increasing temperature,
whereas its depth remains constant.

The DOS presented in Figs. \ref{DOS_smallU} and \ref{DOS_largeU}
are obtained for 20$\times$20 clusters. The question arises whether
the fine structure seen in the DOS is a finite--size effect or
is inherent to the FK model.
The finite--size effects are especially important for small $U$.
This drawback is visible in
Fig. \ref{DOS_smallU}c, where the inset shows the DOS obtained
for 40$\times$40 cluster. In both cases of 20$\times$20 and
40$\times$40 clusters, the overall DOS resembles corresponding DOS
for free electrons on square lattice. However, for the DOS
obtained for 40$\times$40 cluster is much more smooth, and therefore
we attribute the fine structure to the finite size of the cluster.
The DOS presented in Fig. \ref{DOS_smallU}c does not change
with increasing temperature. This indicates that for $U/t=1$
temperature $k_{\rm B}T/t=0.1$ is high enough to treat the electrons
described by the FK Hamiltonian as a free electron gas on a square
lattice with diagonal disorder. Moreover, the interaction $U$ is weak
enough to neglect in this temperature regime the influence
of averaged disorder and the actual DOS is almost the same as for
free electrons on a square lattice.

\begin{figure}[!h]
\centerline{
\includegraphics[width=8.5cm,clip=true]{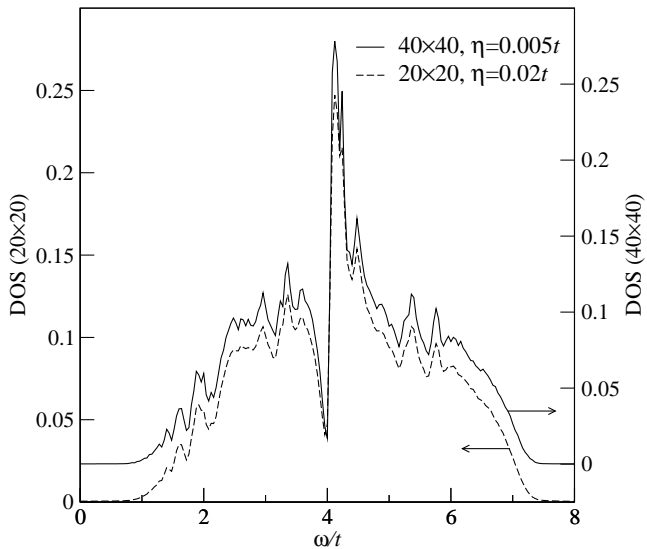}
}
\caption{Comparison of DOS obtained for 20$\times$20 and
40$\times$40 clusters with different delta function broadening.
Note, that the right vertical axis is shifted in order to make
the figure more legible. The presented results have been obtained
for $U/t=8$ at temperature $k_{\rm B}T/t=1$, i.e., the same as in Fig.
\ref{DOS_largeU}c.}
\label{20vs40}
\end{figure}

On the other hand, when $U$ increases, the DOS
becomes much less dependent on the cluster size and the features
that develop in the DOS can be attributed to the FK model itself.
In order to confirm this assumption we compare DOS calculated for
20$\times$20 cluster with delta function broadening $\eta=0.02\:t$
and for 40$\times$40 with $\eta=0.005\:t$. Results are presented
in Fig. \ref{20vs40}.
The smaller broadening used in the case of simulations on the larger
cluster can uncover more details of the actual density of states.
The differences between the DOS obtained for 20$\times$20 and
40$\times$40 clusters with the same broadening $\eta=0.02\:t$
are within the linewidth. The results presented in Fig. \ref{20vs40}
can be directly compared to results obtained using dynamical cluster
approximation (see Fig. 6 in Ref. \onlinecite{DCA}). When the size of
the cluster increases from 1$\times$1 (what corresponds to the dynamical
mean--field
approximation) to 8$\times$8, some features start to develop in the
DOS. The positions of these features are almost the same as in
our MC calculations, however, they are less developed. In particular,
it seems that the DOS at $\omega=\pm U/2$ vanishes in infinite system
and narrow peaks
are located at $|\omega|$ slightly larger than $U/2$.
In Ref. \onlinecite{DCA}, these peaks are attributed to the localization
of electrons in four sites surrounding each site occupied by an ion.
However, these features, as well as the overall density of states, do
not change when the temperature increases. Moreover, the high--temperature
density of states is exactly the same as the one for systems with random
on--site binary disorder. Both the peak and the (pseudo)gap in the
vicinity of $\pm U/2$ have been obtained within many approaches to
disordered binary alloys.\cite{alloy} This indicates that those
peaks are rather connected with the strong interaction between
electrons and randomly distributed ions.

\section{Spectral properties}

The temperature and interaction dependence of the density of states
is connected with the spectral properties of the FK model. Since at
half filling there is no phase separation, the system is translation invariant
and the ionic--configuration--averaged electronic Green function depends
only on one momentum vector
\begin{eqnarray}
&\displaystyle
\sum_{{\bm R}_i} \sum_{{\bm R}_j} \exp\left\{i\left({\bm k}\cdot {\bm R}_i
- {\bm k'}\cdot {\bm R}_j\right)\right\}\langle
{\cal G}\left({\bm R}_i, {\bm R}_j,z \right) \rangle \nonumber \\
&= G\left({\bm k},z\right)\delta\left({\bm k} - {\bm k'}\right).
\end{eqnarray}
Here, ${\cal G}\left({\bm R}_i, {\bm R}_j,z \right)=
\left\{\left[z-{\cal H}({\cal C})\right]^{-1}\right\}_{ij}$ is
the real--space Green function for a given ionic configuration ${\cal C}$
and $\langle\ldots\rangle$ denotes averaging over the configurations.
Then, the spectral function can be determined from the standard formula
\begin{equation}
A({\bm k},\omega) = -\frac{1}{\pi}{\rm Im}\: G\left({\bm k},\omega+i0^+\right).
\end{equation}
\begin{figure*}
\includegraphics[width=18.5cm,clip=true]{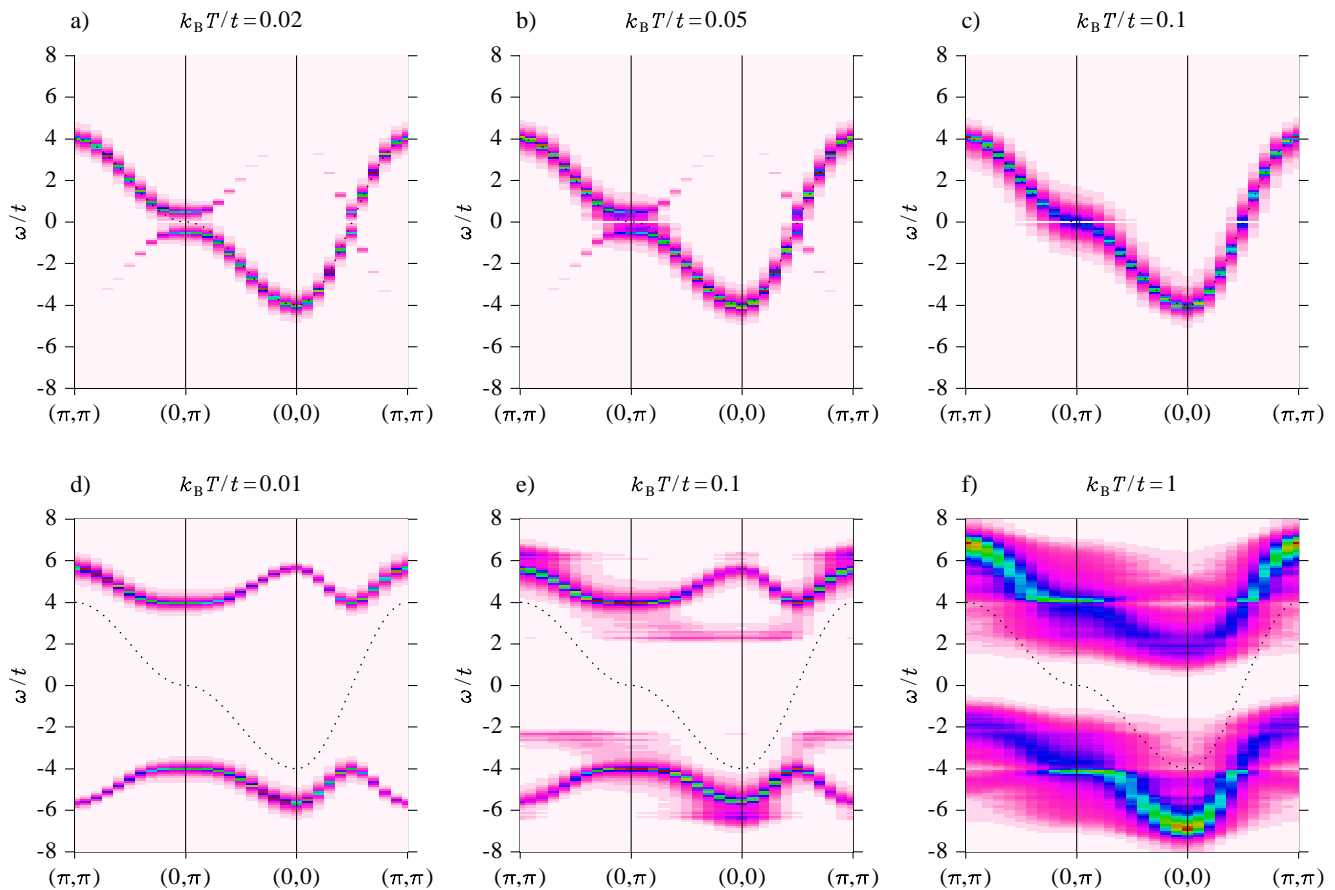}
\caption{(Color online) Spectral functions for $U/t=1$ (upper row) and $U/t=8$ (lower row)
at various temperatures. The dotted line indicates dispersion of free electrons on 
the square lattice.\label{spectr}}
\end{figure*}
Figure \ref{spectr} shows the spectral weights corresponding to the
DOS presented in Figures \ref{DOS_smallU} and \ref{DOS_largeU}.
In the relatively weak interaction regime ($U/t=1$) at low temperature
there is the CDW gap at $(0,\pi)$ and $(\pi/2,\pi/2)$ (Fig. \ref{spectr}a).
The dispersion relation is very accurately described by
$E_{\bm k}=\pm\sqrt{\left(\epsilon_{\bm k}-\mu\right)^2+U^2}$, where
$\epsilon_{\bm k}=-t\left(\cos k_x + \cos k_y\right)$ is the energy of free
electrons on a square lattice. With the increasing temperature
the ionic disorder increases, destroying the checkerboard pattern.
As a result, the spectral peaks gradually broadens up filling the
gap (Fig. \ref{spectr}b). However, due to the weakness of the electron--ion
interaction, this broadening is relatively small. Eventually, at high
temperature the CDW gap disappears and the dispersion relation takes on
the form of $\epsilon_{\bm k}$, that does not change with further increase
of temperature (Fig. \ref{spectr}c).
In the strong interaction regime ($U/t=8$) at
low temperature the dispersion relation can be described by the same
formula, as for $U/t=1$. In this case, however, the stronger interaction
leads to almost equal spectral weights in the lower and upper subband
and the band structure does not resemble that for free electrons any more
(Fig. \ref{spectr}d). When the ionic disorder increases, additional
bands appear, mainly around $(0,0)$ and ($\pi,\pi)$ points (Fig. \ref{spectr}e).
This effect is more visible in the corresponding DOS (Fig. \ref{DOS_largeU}b).
These bands broaden up with farther increase of temperature and most
of the spectral weight is transferred to them (Fig. \ref{spectr}f).
Despite the broadening, at point $(0,\pi)$ there still exist flat parts of
the bands, that are responsible for the peaks in the DOS at energy $\pm U/2$.

There exist some exact results for the mobile particles in the form of sum 
rules and the numerical results can be checked against them.\cite{sum-rule1,sum-rule2} 
They give values of a few lowest spectral moments, defined by 
\begin{equation}
\mu_n({\bf k})=\int_\infty^\infty \omega^n A({\bf k},\omega).
\end{equation}
Their values in the half--filling case are given by very simple 
expressions
\begin{equation}
\mu_0({\bf k})=0, \ \ 
\mu_1({\bf k})=\epsilon_{\bf k},  \ \ 
\mu_2({\bf k})=\epsilon_{\bf k}^2+\frac{U^2}{4}.
\end{equation}
It is interesting that the same results are valid for the Hubbard model\cite{sum-rule3}
and for the FK model in a nonequilibrium state.\cite{sum-rule1}
Since in the proposed approach the FK Hamiltonian for a finite system is 
{\em exactly} diagonalized in each MC step, all these sum rules are 
{\em exactly} satisfied. As a result the sum rules are also satisfied for 
the spectral functions obtained in the whole MC run. However, the results
for the moments are exact only if they are determined directly from the distribution 
of the eigenenergies in each MC step. The other approach, i.e., integrating $\omega^n$
with the spectral function obtained in the whole MC run, leads to a small error
introduced by the broadening of the Dirac $\delta$ functions. The same holds true
for the local moments,\cite{sum-rule1} where $\omega^n$ is integrated with
the density of states.

\section{Summary}
We have presented Monte Carlo analysis of half--filled FK model.
In order to take into account both the classical and fermionic
degrees of freedom, we have derived a modification of the classical
Metropolis algorithm, where the interaction energy is replaced by free
energy, calculated by numerical diagonalization of the FK Hamiltonian for a
given ionic configuration. Such an approach is possible due to
the absence of many--body interactions in the FK Hamiltonian.
Although there is no many--body
term in the Hamiltonian, averaging over ionic configurations
(``annealing'') leads to many--body effects in the FK model.
As a result, the FK model possesses a rich phase diagram. The simulations
presented in this paper concern the case of $\rho_i=\rho_e=0.5$, i.e.,
the case where the ground state is known to be the checkerboard ionic
configuration. Our results illustrate how the system reaches this state
when the temperature is lowered. It is known, that in the strong
interaction limit the FK model maps onto the Ising model, and therefore
the phase transition to the ordered state is of second order.
There is no such a rigorous result for weakly interacting
2D FK model. Our simulations indicate the presence of
a first--order phase transition in this regime. However, the precise
determination of the critical value of the interaction, below which
the phase transition is of first order, requires extensive simulations,
mostly in order to perform the finite--size scaling.

Additionally, we have shown that the order parameter decreases
unusually with increasing temperature in this regime. 
A departure from the Ising--type temperature dependence of the
order parameter has recently been demonstrated also in infinite 
dimensions.\cite{CFJ} 

Performing the MC simulations for various interaction
strengths and at various temperatures, we have constructed the phase
diagram for half--filled FK model. It was shown, that the boundary
line, separating the ordered (checkerboard) and disordered phases,
determined from positions of the peaks in the specific heat and in
the CDW susceptibility coincides with the one obtained from the opening
of a gap in the electronic DOS. The phase diagram has also been determined
in the presence of NNN hopping. It was shown, that there is a point (for
$t'/t \approx 0.7\div 0.8$, depending on the interaction strength) at
which no long range order exists at any finite temperature. For larger
$t'$ the ions at low temperate form horizontal or vertical stripes.

The electronic DOS and spectral functions have been determined for a wide
range of the interaction strength. In the large--$U$ limit, the
distribution of electrons strongly depends on the ionic configuration.
As a result,
the CDW gap develops in the DOS at low temperatures. At high
temperatures, the ions are distributed randomly over the lattice
sites and the electronic DOS is the same as in a free electron system
with a diagonal disorder. On the other hand, in the weak interaction
regime the electrons are hardly affected by the ions, and averaging
over the ionic configurations leads to the same band structure as for free
electrons on a square lattice. We have shown, that the proposed
approach give both the electronic DOS and spectral functions in a good
agreement with that obtained with use of much more elaborated DCA method,
whereas the computational effort is much lower.

The present paper demonstrated how the MC approach can be used
to describe the formation of the checkerboard order in the
half--filled FK model. It is also possible to use this method
to analyze the FK model away from half--filling, where inhomogeneous
states are expected to minimize the free energy at low enough
temperatures. The MC method would be particularly useful
to describe phases with irregular ionic configurations, e.g.,
phase separation, which are difficult to analyze analytically.
Preliminary result in this area are presented in Ref. \onlinecite{pss}.
Another area where the Monte Carlo study of the Falicov--Kimball may be
particularly useful, includes nonbipartite lattices. In this case, the
checkerboard pattern cannot be formed even at half--filling, and the model
is expected to exhibit a strong frustration. This issue is under current
investigation.\cite{sces05}

\acknowledgments
The authors acknowledge stimulating discussions with Jim Freericks,
Romuald Lema\'nski, Marcin Mierzejewski, and Stanis{\l}aw Robaszkiewicz.
The authors also acknowledges support from the Polish Ministry of Education 
and Science under Grant No. 1 P03B 071 30.

\end{document}